\documentclass[twocolumn,eqsecnum,aps,epsfig]{revtex4}
\usepackage{epsfig}

\newcommand{\beq}{\begin{eqnarray}}
\newcommand{\eeq}{\end{eqnarray}}

\begin{document}

\title {Structure of $A=7$ iso-triplet $\Lambda$ hypernuclei
studied with the four-body cluster model}

\author{E.\ Hiyama}

\address{Nishina Center for
Accelerator-Based Science,
Institute for Physical and Chemical
Research (RIKEN), Wako, Saitama,
351-0198,Japan}

\author{Y. Yamamoto}

\address{Physics Section, Tsuru University, Tsuru, Yamanashi 402-8555, Japan}

\author{T. Motoba}

\address{Laboratory of Physics, Osaka Electro-Comm.
University, Neyagawa 572-8530, Japan}

\author{M. Kamimura}

\address{Department of Physics, Kyushu University,
Fukuoka,812-8581,Japan}

%

\begin{abstract}
The structure of the $T=1$ iso-triplet hypernuclei,
$^7_{\Lambda}$He, $^7_{\Lambda}$Li and $^7_{\Lambda}$Be within the
framework of an $\alpha +\Lambda +N+N$ four-body cluster model is studied.
Interactions between the constituent subunits are determined so as to
reproduce reasonably well the observed low-energy properties of the $\alpha N$,
 $\alpha \Lambda$, $\alpha NN$ and $\alpha \Lambda N$ subsystems.
 Furthermore, the two-body $\Lambda N$ interaction
 is adjusted so as to reproduce
 the $0^+$-$1^+$ splitting of $^4_{\Lambda}$H.
 Also a phenomenological $\Lambda N$ charge symmetry breaking(CSB) interaction
 is introduced.
 The $\Lambda$ binding energy of the ground state in $^7_{\Lambda}$He
 is predicted to be 5.16(5.36) MeV with(without) the CSB interaction.
 The calculated energy splittings of the $3/2^+$-$5/2^+$  states in
 $^7_{\Lambda}$He and $^7_{\Lambda}$Li are around 0.1 MeV.
 We point out that there is a  three-layer structure of the
 matter distribution, $\alpha$ particle, $\Lambda$ skin,
  proton or neutron halo, in the $^7_{\Lambda}{\rm He}(J=5/2^+)$,
  $^7_{\Lambda}{\rm Li}(J=5/2^+)$ and $^7_{\Lambda}{\rm Be}(J=1/2^+)$ states.
\end{abstract}

\maketitle

\section{Introduction}

A new stage in hypernuclear physics has been opened
by the $\gamma$-ray spectroscopy for $\Lambda$ hypernuclei,
where level structures of the order of keV are revealed systematically.
In order to extract valuable information on hypernuclear structure
and underlying $\Lambda N$ interactions from these extremely
precise data; it is therefore indispensable
to utilize accurate models for the many-body wave functions.

Our special concern in this work is the structure of a multiplet of
$\Lambda$ hypernuclei specified by an isospin $T$,
which have provided us with many interesting subjects so far.
For example, in the case of the $T=1$ multiplet with mass number $A=7$,
$^7_{\Lambda}$He, $^7_{\Lambda}$Li and $^7_{\Lambda}$Be,
their core nuclei are neutron or proton halo nuclei.
When a $\Lambda$ particle is added to the core nuclei,
$^6$He, $^6$Li($T=1$) and $^6$Be,
the resultant hypernuclear systems become more stable against
neutron or proton emission.
Hereafter, $T=1$ excited states of
$^6$Li and $^7_{\Lambda}$Li are denoted as
$^6$Li$^*$ and $^7_{\Lambda}$Li$^*$.
This stabilization is caused by the so-called
"gluelike" role of the $\Lambda$ \cite{Motoba85}.
Thanks to the role of $\Lambda$ particle, we can expect
an interesting possibility that neutron (proton) drip lines
in $\Lambda$ hypernuclei are extended far away from those
in ordinary nuclear systems.

In the past, the level structures in $^7_{\Lambda}$He,
$^7_{\Lambda}$Li ($T=1)$ and $^7_{\Lambda}$Be were studied with the
three-body $^5_{\Lambda}$He$+N+N$ model \cite{Hiyama96}, where
only the even-state $\Lambda N$ interaction was used.
In Ref. \cite{Hiyama96},
we pointed out that there appear halo or skin structures
in the ground state or some excited states of
these hypernuclei.
Recently, the experimental energy of the $T=1$ $J=1/2^+$ state
of $^7_{\Lambda}$Li has been observed through
the high-resolution $\gamma$-ray experiment
\cite{Tamura00}.
Furthermore, it is proposed to produce $^7_{\Lambda}$He
by $(e,e'K^+)$ at JLAB.
One aim in the present work is
to discuss halo or skin structure in the extended  framework
of an $\alpha +\Lambda +N+N$ four-body model.

Another interesting subject to discuss the spin-doublet state,
$5/2^+$-$3/2^+$  in $^7_{\Lambda}$He
and $^7_{\Lambda}$Li ($T=1$).
It is considered that
these excited $5/2^+$-$3/2^+$ doublets are related  intimately to the
spin-dependent potentials of the $\Lambda N$ interaction.
Therefore, it is important to discuss these splitting energies to
determine the spin-dependent parts of the $\Lambda N$ interaction.

In our previous work \cite{Hiyama06},
the spin-doublet structures of $^7_\Lambda$Li
in $T=0$ states and the underlying spin-dependent interactions were
investigated successfully in the $\alpha pn\Lambda$ four-body cluster model.
Here, the $\alpha p$ and $\alpha n$ interactions were chosen so as to
reproduce the corresponding phase shifts, and the $\Lambda \alpha$
interaction was done so as to reproduce the experimental value of
$B_\Lambda(^5_\Lambda$He), and the $\Lambda N$ spin-spin
(spin-orbit) interaction was fitted so as to be consistent with
the $0^+$-$1^+$ ($5/2^+$-$3/2^+$ ), spin-doublet energy separation
in $^4_\Lambda$H ($^9_\Lambda$Be).
In the present work, our four-body analyses for $^7_\Lambda$Li ($T=0$)
is extended straightforwardly to the $T=1$ multiplet
($^7_\Lambda$He, $^7_\Lambda$Li$^*$, $^7_\Lambda$Be),
where an asterisk stands for the $T=1$  excited states.

An important subject related to the isospin multiplet of
$\Lambda$ hypernuclei is the
charge symmetry breaking (CSB) components in $\Lambda N$ interactions.
The most reliable evidence for the CSB interaction appears in the
$\Lambda$ binding energies $B_\Lambda$ of the $A=4$ members with
$T=1/2$ ($^4_\Lambda$He and $^4_\Lambda$H). Then, the CSB effects
are attributed to the differences
$\Delta_{CSB}=B_\Lambda(^4_\Lambda$He$)-B_\Lambda(^4_\Lambda$H),
the experimental values of which are $0.35\pm0.06$ MeV and
$0.24\pm0.06$ MeV for the ground ($0^+$) and excited ($1^+$)
states, respectively.
There exist mirror hypernuclei in the $p$-shell region
such as the $T=1$ multiplet with $A=7$ ($^7_\Lambda$He, $^7_\Lambda$Li$^*$,
$^7_\Lambda$Be), $T=1/2$ multiplet with $A=8$
($^8_\Lambda$Li, $^8_\Lambda$Be), $T=1/2$ multiplet with $A=10$
($^{10}_\Lambda$Be, $^{10}_\Lambda$B), and so on.
Historically, some authors mentioned CSB effects in these $p$-shell
$\Lambda$ hypernuclei \cite{Gal77,Gibson95}.
 However, there is no microscopic calculation of these
hypernuclei taking account of the CSB interaction.

It is well known that the experimental values $\Delta_{CSB}$ can be
fitted phenomenologically by an effective spin-independent CSB
interaction. On the other hand, in the case of a meson-theoretical model
an OPE-type CSB potential is derived through
a $\Lambda - \Sigma^0$ mixing effect,
where the triplet CSB interaction is much stronger than the singlet one
due to the tensor-force contribution. This feature is in strong
disagreement with that in the phenomenological force which is
almost spin-independent.
This difference between triplet and singlet CSB interactions
appears in the elaborate four-body calculations for
$^4_\Lambda$He and $^4_\Lambda$H with use
of the Nijmegen soft core model (NSC97e model) \cite{Nogga02},
in which the CSB components are generated
by the mass difference within the $\Sigma$-multiplet mixed in
$\Lambda$ states and the $\Lambda - \Sigma^0$ mixing effect.

Because the origin of the CSB interaction is not yet settled,
we treat it phenomenologically in the present study:
Similarly to Ref.\cite{Bodmer84}, the CSB interaction is determined so as to
reproduce the values of $\Delta_{CSB}$ obtained from
the $\Lambda$ binding energies of $^4_\Lambda$H and $^4_\Lambda$He.
Then, the $T=1$ triplet hypernuclei with $A=7$
($^7_\Lambda$He, $^7_\Lambda$Li$^*$, $^7_\Lambda$Be) are studied
with use of this CSB interaction in the four-body cluster model.
Additionally, the CSB effects in the $T=1/2$ doublet hypernuclei
with $A=8$ are investigated within the $\alpha t \Lambda$
and $\alpha ^3{\rm He} \Lambda$ cluster models for $^8_\Lambda$Be
and $^8_\Lambda$Li, respectively.

In this work,
we study $A=7$ hypernuclei within the framework of
an $\alpha +\Lambda +N+N$ four-body model
so as to take account of the full correlations among
all the constituent baryons.
Two-body interactions
among constituent particles are
chosen so as to reproduce all the existing
binding energies of the sub-systems ($\alpha N, \alpha \Lambda N,
\alpha \Lambda$, and so on).
This feature is important in the analysis of the
energy levels of these hypernuclei.
Our analysis is performed systematically for ground and  excited states
of $\alpha \Lambda NN$ systems
with no more adjustable parameters in this stage,
so that these predictions offer important guidance for the
interpretation of the upcoming hypernucleus experiments such as
the $^7$Li$(e,e'K^+)$ $^7_{\Lambda}$He reaction at
Thomas Jefferson National Accelerator Facility (JLAB).

In Sec. II, the microscopic $\alpha \Lambda NN$ and $NNN \Lambda$ four-body
calculation method is described.
In Sec.III, the interactions are explained.
The calculated results and the discussion are presented in Sec.IV.
Sec. V is devoted to the discussion on the charge symmetry
 breaking effects obtained for the A=7 and 8 systems. The summary
 is given in Sec. VI.

\section{Four-body cluster model and method}

The models employed in this paper are
the same as those in our previous work \cite{Hiyama06}.
Namely, we employ the $\alpha +\Lambda +N+N$ model
for the $A=7$ hypernuclei (Fig.1) and the $\alpha +N+N$
model for the $A=6$ nuclei (Fig.3 in Ref.\cite{Hiyama06}),
where all the rearrangement channels are taken into account.
%
\begin{figure}[htb]
\begin{center}
\epsfig{file=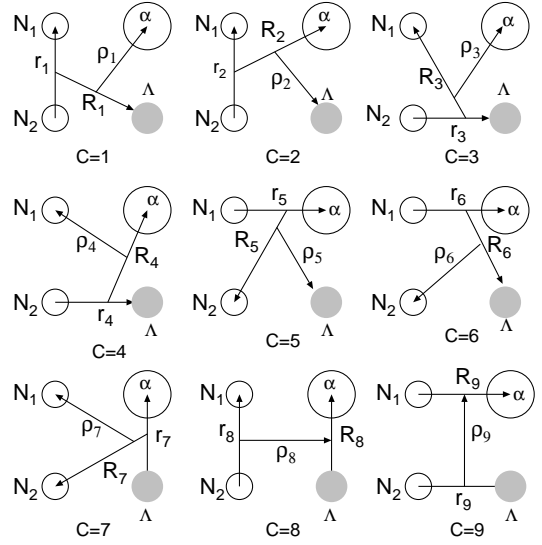,scale=0.35}
\end{center}
\caption{Jacobi coordinates for all
the rearrangement channels ($c=1 \sim 9$)
of the $\alpha +\Lambda +N_1+N_2$ four-body system.
Two nucleons are to be antisymmetrized.}
\label{fig:jacobi7}
\end{figure}
%
The Schr\"{o}dinger equation
is given by
\begin{eqnarray}
&& \qquad \qquad
( H - E ) \, \Psi_{JM,TT_z}(^7_{\Lambda}{\rm Z})  = 0 ,\\
\label{eq:schr7}
&& \!\!\!\!\!\!\!\!\!\! \!\!\!\!\!\!  H=T+ \!V_{N_1 N_2}
      + \!\sum_{i=1}^2 ( V_{\Lambda N_i}
     \! + \!V_{\alpha N_i})
      + \!V_{\alpha \Lambda}
          +\!V_{\rm Pauli},
\label{eq:hamil7}
\end{eqnarray}
where $V_{\alpha N_i}$ is the interaction between
the $\alpha$ particle and $i$-th nucleon and
$V_{\alpha \Lambda}$ is the $\alpha \Lambda$ interaction,
which are explained in the next section.
The Pauli principle between
the  $\alpha$ particle and the two nucleons
is taken into account by the
Pauli projection operator
$V_{\rm Pauli}$, which is the same as
in Ref.\cite{Hiyama06}.
The total wave function is described
as a sum of amplitudes of all the rearrangement channels
shown in
Fig.~\ref{fig:jacobi7} in the $LS$ coupling scheme:
\begin{eqnarray}
   \Psi&& \!\!\!\!\!\!\!_{JM, TT_z}(^7_{\Lambda}{\rm Z})
      \nonumber \\
&& =  \sum_{c=1}^{9} \:
      \sum_{nl, NL, \nu \lambda}
       \sum_{IK} \sum_{sS}
       C^{(c)}_{nl,NL,\nu\lambda, IK, sS}\: \Phi(\alpha) \nonumber  \\
      &&  \times    {\cal A}
      \Big\{ \Big[
        \big[ \, [ \phi^{(c)}_{nl}({\bf r}_c)
         \psi^{(c)}_{NL}({\bf R}_c)]_I
      \,  \xi^{(c)}_{\nu\lambda} (\mbox{\boldmath $\rho$}_c)
        \big]_{K} \nonumber  \\
     && \times
       \big[ \,
     [ \chi_{\frac{1}{2}}(N_1)
       \chi_{\frac{1}{2}}(N_2)
       ]_s \chi_{\frac{1}{2}}(\Lambda) \big]_S
           \Big]_{JM} \nonumber \\
     && \times     \big[\, \eta_{\frac{1}{2}}(N_1)
       \eta_{\frac{1}{2}}(N_2) \,
       \big]_{T T_z} \Big\}    \;  ,
\label{eq:he7lwf}
\end{eqnarray}
where the notations are the same as  in Ref. \cite{Hiyama06}.
Also, the definitions of the Gaussian basis functions
and the Gaussian ranges are the same as those in the case of
the $A=4$ hypernuclei.

  The eigenenergy $E$  in Eq.~(\ref{eq:schr7}) and the
$C$ coefficients in Eq.~(\ref{eq:he7lwf}) are determined by
the Rayleigh-Ritz variational method.
The angular momentum space of $l, L, \lambda \leq 2$
was found to
be sufficient to obtain good convergence of the
calculated results as described below.

\section{Interactions}

\label{interaction}

\subsection{Charge symmetry parts}

We recapitulate here the charge symmetric  parts of the
$V_{N\alpha}$, $V_{NN}$, $V_{\alpha \Lambda}$, and $V_{\Lambda N}$
interactions employed in our
$\alpha NN \Lambda$ systems \cite{Hiyama06}.

For $V_{N \alpha}$, we employ the effective potential
proposed in Ref.\cite{Kanada79}, which is designed so as to reproduce
well the low-lying states and low-energy scattering phase
shifts of the $\alpha n$ system.
The Pauli principle between nucleons belonging to the
$\alpha$ and the valence nucleon is taken into account
by the orthogonality condition model (OCM) \cite{Saito69}.
As for  the $NN$ interaction $V_{NN}$,
we use the AV8 \cite{Pudliner97} potential,
derived from the AV18 \cite{Wiringa95}
by neglecting the $(L \cdot S)$ term.

The interaction $V_{\alpha \Lambda}$ is obtained by folding the
$\Lambda N$ G-matrix interaction derived from
the Nijmegen model F(NF)~\cite{NDF}
into the density of the $\alpha$ cluster~\cite{Hiyama97}, its strength
being adjusted so as to reproduce the experimental value of
$B_\Lambda(^5_\Lambda$He).

For $V_{\Lambda N}$, we employ effective single-channel
interactions simulating the basic features of the
Nijmegen model NSC97f~\cite{NSC97}, where
the $\Lambda N$-$\Sigma N$ coupling effects are renormalized
into $\Lambda N-\Lambda N$ parts:
We use three-range Gaussian potentials  so as to reproduce
the $\Lambda N$ scattering phase shifts calculated
from the NSC97f, and then
their second-range strengths in $^3E$ and $^1E$ states are
adjusted so that calculated energies of
$0^+$-$1^+$ doublet state in the $NNN\Lambda$
four-body calculation reproduce the observed
splittings of $^4_\Lambda$H.
Furthermore, the spin-spin parts in the odd states are tuned
to get the experimental values of
the splitting energies of $^7_\Lambda$Li.
The symmetric LS (SLS) and anti-symmetric LS (ALS)
parts in $V_{\Lambda N}$ are chosen so as to
be consistent with the $^9_\Lambda$Be data as follows:
The SLS and ALS parts derived from NSC97f with the G-matrix
procedure are represented in the two-range form, and then
the ALS part is strengthened so as to reproduce
the measured $5/2^+$-$3/2^+$ splitting energy with the
$2\alpha + \Lambda$ cluster model~\cite{Hiyama00}.
The parameters in the $\Lambda N$ interactions are given in
\begin{eqnarray}
&& V_{\Lambda N}(r)=  \\ \nonumber
&&\sum^3_{i=1} \frac{1+P_r}{2}(v_{0}^{i,\rm even} +
\mbox{\boldmath $\sigma$}_\Lambda \cdot
\mbox{\boldmath $\sigma$}_N
v_{\sigma_\Lambda \cdot
\sigma_N}^{i,\rm even})
\, e \:^{-\beta^i_{\Lambda N}r^2}    \\ \nonumber
&+& \frac{1-P_r}{2}(v_{0}^{i,\rm odd} +
\mbox{\boldmath $\sigma$}_\Lambda \cdot
\mbox{\boldmath $\sigma$}_N
v_{\sigma_\Lambda \cdot
\sigma_N}^{i,\rm odd})
\, e \:^{-{\beta^i_{\Lambda N}}r^2}   ,
\label{eq:lambdan}
\end{eqnarray}
and listed in Table I(a).

The calculated energies of the $0^+$ states in
$^6$He and $^6$Li$^*$ are $-0.59$ MeV and unbound with
the respect to the $\alpha +N+N$ three-body breakup threshold,
which are less bound than the observed values,
$-0.98$ MeV in $^6$He and $-0.14$ MeV in $^6$Li.
 Considering that
it is of vital importance in our cluster model to reproduce
accurately the binging energy of all subcluster systems,
we introduce an effective three-body $\alpha NN$ interaction
phenomenologically, the form of which is assumed as

\begin{equation}
V_{\alpha NN}(r_1,r_2)=
 \sum^2_{i=1} v_i e \:^{-\beta^i r_1^2-\beta^ir_2^2},
\end{equation}
where $r_1$ and $r_2$ are Jacobian coordinates for $C=1$ and 2 in Fig. 3 of
Ref.\cite{Hiyama96}.

This interaction includes four parameters ($\beta^i$, $v_i$),
which cannot be determined completely by the two binding energies of
$^6$He and $^6$Li$^*$ only. Then, the condition to reproduce
the experimental value of $^7_\Lambda$Li$^*$ is found to give
a strong constraint for the parameters.
The determined values of parameters are ($\beta^1, v_1$)=
(0.444 fm$^{-2}$, 244.8 MeV),
($\beta^2, v_2$)=(0.128fm$^{-2}$, $-20.4$ MeV).

\subsection{Charge symmetry breaking interaction}

\label{sec:CSB}

It is out of our scope in this work to explore the origin of
the CSB interaction. We assume  here the CSB interaction
with an one-range Gaussian form
\begin{eqnarray}
&& V_{\Lambda N}^{\rm CSB}(r)= \\ \nonumber
&&-\frac{\tau_z}{2}\Big[\frac{1+P_r}{2}(v_{0}^{\rm even,CSB} +
\mbox{\boldmath $\sigma$}_\Lambda \cdot
\mbox{\boldmath $\sigma$}_N
v_{\sigma_\Lambda \cdot
\sigma_N}^{\rm even, CSB})
\, e\:^{-\beta_{\rm even}\: r^2}\   \\ \nonumber
&+& \frac{1-P_r}{2}(v_{0}^{\rm odd,CSB} +
\mbox{\boldmath $\sigma$}_\Lambda \cdot
\mbox{\boldmath $\sigma$}_N
v_{\sigma_\Lambda \cdot
\sigma_N}^{\rm odd,CSB})
\, e \:^{-\beta_{\rm odd} \:r^2}\ \Big]  ,
\end{eqnarray}
which includes spin-independent and spin-spin parts.
In the cases of the four-body calculations of $^4_{\Lambda}$H
($nnp\Lambda$) and $^4_{\Lambda}$He ($npp\Lambda$),
the contributions of the odd-state interactions are
negligibly small and their strengths cannot be determined:
We take $v_0^{\rm odd,CSB}=0$, and
$v_{\sigma_\Lambda \cdot
\sigma_N}^{\rm odd,CSB}=0$.
The range parameter, $\beta_{\rm even}$ is taken to be 1.0 fm$^{-2}$.
The parameters $v_0^{\rm even}$ and $v_{\sigma \sigma}^{\rm even}$
are determined phenomenologically so as to reproduce the values of
$\Delta_{CSB}$ derived from the $\Lambda$ binding energies of
$0^+$ and $1^+$ states in the
four-body calculation of $^4_\Lambda$H ($^4_\Lambda$He).
Then, we obtain $v_0^{\rm even,CSB}=8.0$ MeV and
$v_{\sigma \sigma}^{\rm even,CSB}$=0.7 MeV.
The calculated $B_{\Lambda}$ of $0^+$ and $1^+$ states in
$^4_{\Lambda}$H are 1.99 MeV and 0.98 MeV, respectively.
Those in $^4_{\Lambda}$He are 2.35 MeV and 1.17 MeV, respectively.
In these calculations including the CSB interactions,
the parameters in the CS parts are slightly modified from those in
Table Ia for fine fitting of the experimental $B_\Lambda$ values.
In Table I(a), the modified values of parameters are given in the parentheses.

In order to extract the information about the odd-state part of CSB,
it is necessary to study iso-multiplet hypernuclei in the $p$-shell region.
A suitable system for such a study is $^7_{\Lambda}$He,
in which the core nucleus $^6$He is in a bound state.
(On the contrary, valence protons in $^6$Be are unbound.)
Our four-body calculation for this system has to be powerful
to extract the accurate information.
Though there is no data about $^7_{\Lambda}$He at present,
the coming experiments at JLAB will give us valuable data
for our analyses.

Another example in the $p$-shell region is iso-doublet hypernuclei
$^8_{\Lambda}$Li and $^8_{\Lambda}$Be, whose experimental values of
$B_\Lambda$ are obtained in emulsion.
Then, it is interesting to see the contribution of the CSB interaction
to the $B_\Lambda$ values of these hypernuclei.
For applications to these nuclei, we used $\Lambda$-$t$ and
$\Lambda$-$^3{\rm He}$ potential for CS part defined by
\begin{eqnarray}
&& V_{\Lambda x}({\bf r,r'})=  \\ \nonumber
&&\sum^3_{i=1}
\frac{1}{2} \Big[
(V_{0}^{i,{\rm even}} +
{\bf s}_\Lambda \cdot
{\bf s}_x
V_{S}^{i,{\rm even}})
\, e\:^{-\mu^{i}r^2}\:
\delta({\bf r-r'})  \\ \nonumber
&+&
(U_{0}^{i,{\rm even}} +
{\bf s}_\Lambda \cdot
{\bf s}_x
U_{S}^{i,{\rm even}})
\, e\:^{-\gamma^i({\bf r+r'})^2-\delta^i
({\bf r-r'})^2}\Big]  \\ \nonumber
&+&
\frac{1}{2} \Big[
(V_{0}^{i,{\rm odd}} +
{\bf s}_\Lambda \cdot
{\bf s}_x
V_{S}^{i,{\rm odd}})
\, e\:^{-\mu^{i}r^2}\:
\delta({\bf r-r'})  \\ \nonumber
&+&
(U_{0}^{i,{\rm odd}} +
{\bf s}_\Lambda \cdot
{\bf s}_x
U_{S}^{i,{\rm odd}})
\, e\:^{-\gamma^i({\bf r+r'})^2-\delta^i
({\bf r-r'})^2}\Big]  \quad,
\label{eq:tx}
\end{eqnarray}
where $x$ denotes t or $^3$He.
The parameters are listed in Table I(b).
The CSB part for $\Lambda$-$t$ and $\Lambda$-$^3$He is
given by
\begin{eqnarray}
&& V_{\Lambda x}^{\rm CSB}({\bf r,r'})=  \\ \nonumber
&&\frac{1}{2} \Big[
(V_{0}^{\rm even,CSB} +
{\bf s}_\Lambda \cdot
{\bf s}_x
V_{S}^{\rm even,CSB})
\, e\:^{-\mu_{\rm even}r^2}\:
\delta({\bf r-r'})  \\ \nonumber
&+&
(U_{0}^{\rm even,CSB} +
{\bf s}_\Lambda \cdot
{\bf s}_x
U_{S}^{\rm even,CSB})
\, e\:^{-\gamma_{\rm even}({\bf r+r'})^2-\delta_{\rm even}
({\bf r-r'})^2}\Big]  \\ \nonumber
&+&
\frac{1}{2} \Big[
(V_{0}^{\rm odd,CSB} +
{\bf s}_\Lambda \cdot
{\bf s}_x
V_{S}^{\rm odd,CSB})
\, e\:^{-\mu_{\rm odd}r^2}\:
\delta({\bf r-r'})  \\ \nonumber
&+&
(U_{0}^{\rm odd,CSB} +
{\bf s}_\Lambda \cdot
{\bf s}_x
U_{S}^{\rm odd,CSB})
\, e\:^{-\gamma_{\rm odd}({\bf r+r'})^2-\delta_{\rm odd}
({\bf r-r'})^2}\Big]  \quad.
\label{eq:csb}
\end{eqnarray}
The parameters for even-state are adjusted
so as to reproduce the data within
the $\Lambda$-$t$ and $\Lambda$- $^3$He cluster models for
$^4_\Lambda$H and $^4_\Lambda$He, respectively.
The parameters are
$V_{0}^{\rm even, CSB}=0.38$ MeV,
$V_{S}^{\rm even, CSB}=-0.03$ MeV,
$\mu_{\rm even}=0.06$ fm$^{-2}$,
$U_0^{\rm even, CSB}=0.08$ MeV,
$U_S^{\rm even, CSB}=-0.006$ MeV,
$\gamma_{\rm even}=0.203$ fm$^{-2}$
and $\delta_{\rm even}=0.679$fm$^{-2}$ for $^8_{\Lambda}$Li,
and the same value with the opposite sign for
$^8_{\Lambda}$Be.
As explained later, also the odd-state CSB interaction is introduced
phenomenologically so as to reproduce the $B_\Lambda$ values of
$^8_\Lambda$Li and $^8_\Lambda$Be.
The determined parameters are
$V_{0}^{\rm odd, CSB}=-0.93$ MeV,
$V_{S}^{\rm odd, CSB}=-0.12$ MeV,
$\mu_{\rm odd}=0.223$ fm$^{-2}$,
$U_0^{\rm odd, CSB}=-0.14$ MeV,
$U_S^{\rm odd, CSB}=-0.095$ MeV,
$\gamma_{\rm odd}=0.203$ fm$^{-2}$
and $\delta_{\rm odd}=0.254$fm$^{-2}$ for $^8_{\Lambda}$Li
and the same value with the opposite sigh for
$^8_{\Lambda}$Be.
It is notable here that the odd-state CSB is of far longer range than
the even-state one.

\begin{table}
\caption{
(a)Parameters of the $\Lambda N$ interaction without CSB
interaction defined in Eq.(\ref{eq:lambdan}).
Range parameters are in fm$^{-2}$ and the strengths are in MeV.
The numbers in parentheses are even-state strengths adjusted so as to
reproduce the observed spin doublet state both in
$^4_{\Lambda}$H and $^4_{\Lambda}$He with CSB interaction.
(b) Parameters of the $t(^3{\rm He})\Lambda$ interaction without CSB
interaction defined in Eq.(\ref{eq:tx}).
The numbers in parentheses are adjusted even-state strengths so
as to reproduce the observed spin doublet state both in
$^4_{\Lambda}$H and $^4_{\Lambda}$He with CSB interaction
within the framework of $t(^3{\rm He})\Lambda$ two-body model.
}
\label{tab:halo_energy_rms_A=7}
\begin{tabular}{ccccccc}
\hline   \hline
&  \multicolumn{3}{c}{(a)$\Lambda N$ interaction}  \\
$i$ & &1 &2  &3  \\
$\beta^i_{\Lambda N}$  & &0.391  &1.5625  &8.163 \\
\hline
$v_0^{i,\rm even}$ & &$-3.94$  &\quad $-126.1(-126.4)$  &\quad1943 \\
$v_{\sigma\sigma}^{i,\rm even}$ & &\quad$-0.003$  &\quad17.5($18.0)$
 &\quad$-374.1$\\
$v_0^{i,\rm odd}$ & &$-1.43$  &\quad $72.8$  &\quad3247 \\
$v_{\sigma\sigma}^{i,\rm odd}$ & &\quad$-0.26$  &\quad \quad
$-61.35$  &\quad$-270.9$ \\
\hline
& \multicolumn{3}{c}{(b) $t(^3{\rm He})\Lambda$ interaction} \\
$\mu^i$  & &0.2874  &0.4903  &0.6759 \\
$V^{i,{\rm even}}_0$ & &$-16.37(-16.39)$  &$-145.7(-146.1)$ &172.02(172.01) \\
$V^{i,{\rm even}}_S$ & &0.234(0.229) &16.76(16.76)  &$-20.55(-20.53)$ \\
$V^{i,{\rm odd}}_0$ & &$-11.94(-11.98)$  &$-70.27(-70.36)$ &679.8(678.4) \\
$V^{i,{\rm odd}}_S$ & &4.525(4.537) &5.248(5.237)  &$-233.3(-233.8)$ \\
$\gamma^i$  & &0.2033  &0.2033  &0.2033 \\
$\delta^i$ & &0.3383 &0.8234  &2.521  \\
$U^{i,{\rm even}}_0$
 & &$-1.995(-1.998)$  &$-36.898(-36.99)$ &$156.9(156.9)$ \\
$U^{i,{\rm even}}_S$ & &$0.029(0.028)$ &4.246(4.242)  &$-18.75(-18.73)$  \\
$U^{i,{\rm odd}}_0$
 & &$-1.455(-1.457)$  &$-17.791(-17.814)$ &$620.2(618.9)$ \\
$U^{i,{\rm odd}}_S$ & &$0.552(0.553)$ &1.329(1.326)  &$-212.8(-213.3)$  \\
\hline \hline
\end{tabular}
\end{table}

\section{Results}

First, let us show the level structures of the $T=1$ states
calculated with the $\alpha +\Lambda +N+N$ four-body model
using the same $\Lambda N$ interaction in Ref.\cite{Hiyama06}.
We calculated the bound states in those $\Lambda$ hypernuclei.

\begin{figure*}[htb]
\begin{center}
\epsfig{file=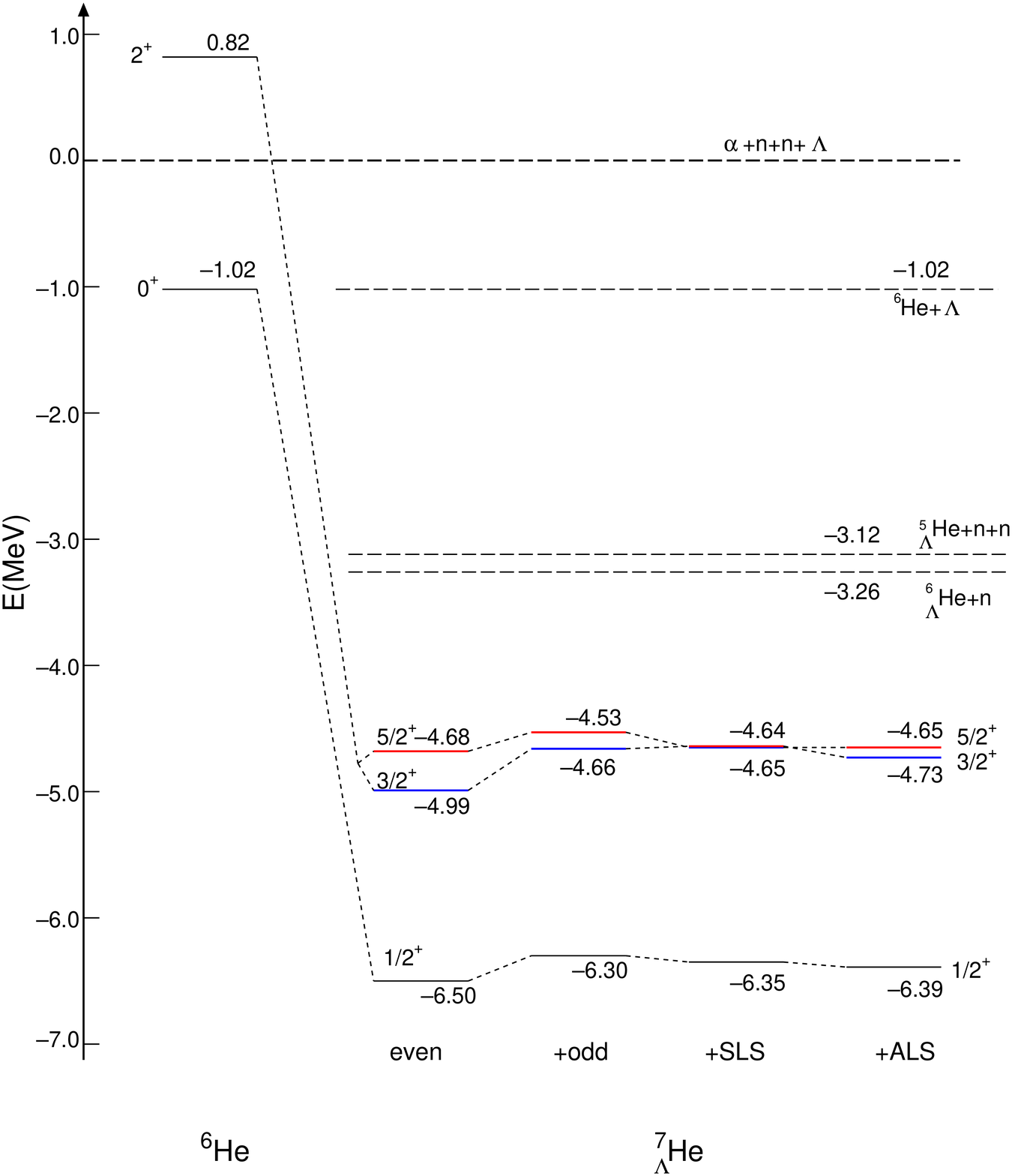,scale=0.35}
\label{fig:he7l}
\caption{(color online).
Calculated energy levels of $^6$He and $^7_{\Lambda}$He.
The charge symmetry breaking potential is not included in
 $^7_{\Lambda}$He.
 The level energies are measured from
 the particle breakup threshold.}
\end{center}
\end{figure*}

\begin{figure*}[htb]
\begin{center}
\epsfig{file=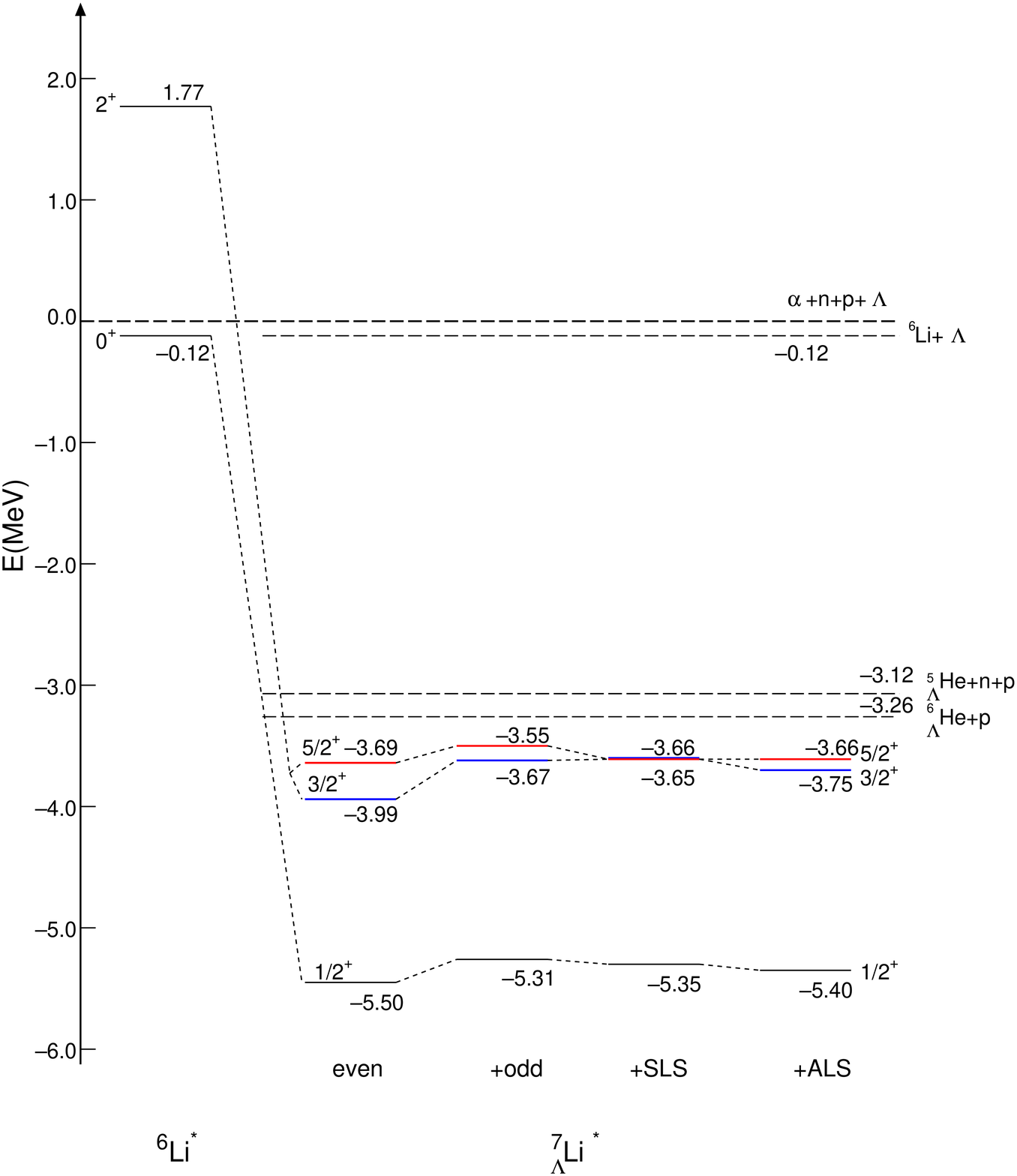,scale=0.35}
\label{fig:li7l}
\caption{(color online).
Calculated energy levels of $^6$Li$^*$ and $^7_{\Lambda}$Li$^*$.
The charge symmetry breaking potential is not included in
 $^7_{\Lambda}$Li$^*$. The level energies are measured from
 the particle breakup threshold.}
\end{center}
\end{figure*}

\begin{figure*}[htb]
\begin{center}
\epsfig{file=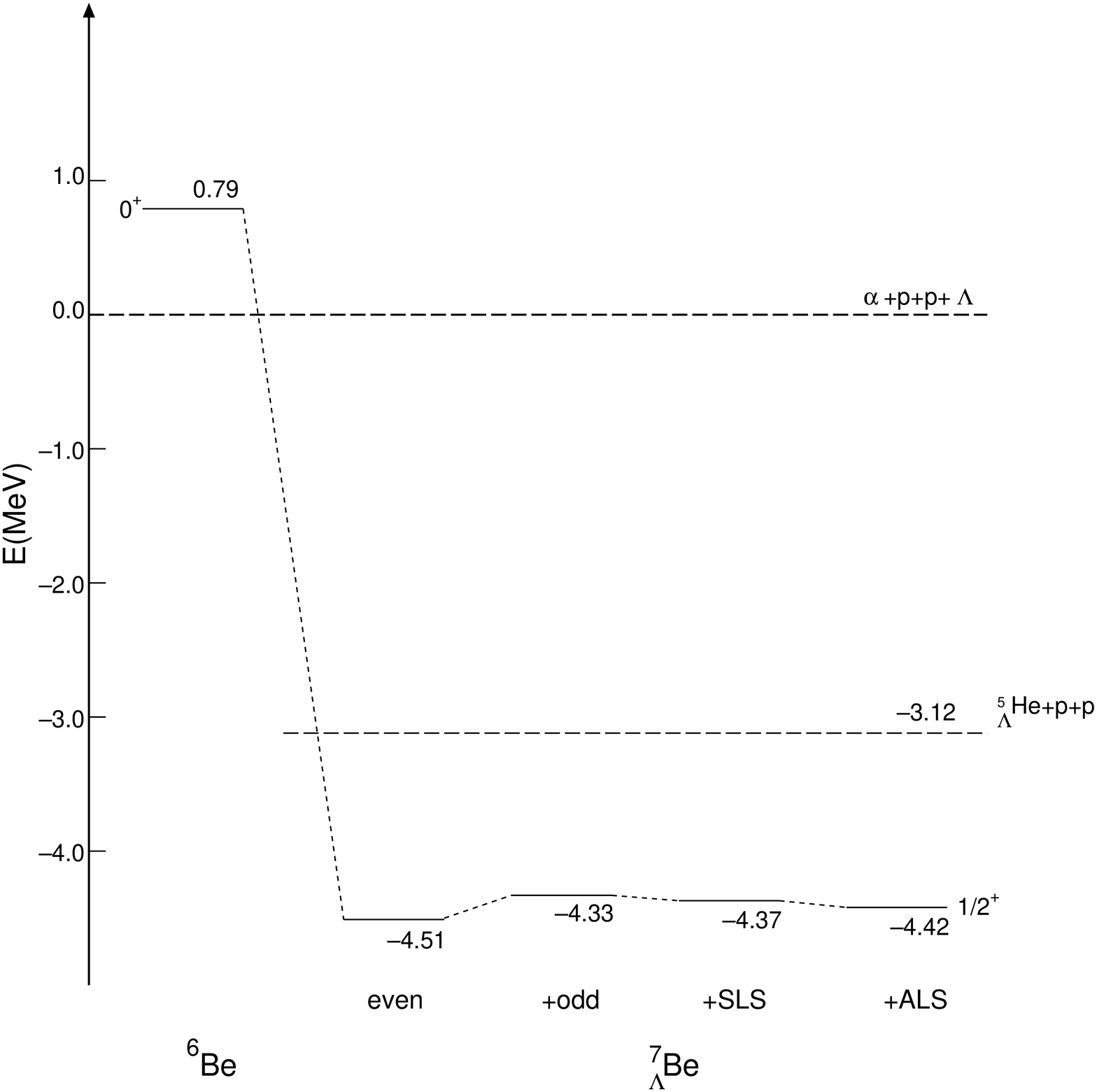,scale=0.35}
\label{fig:be7l}
\caption{Calculated energy levels of $^6$Be and $^7_{\Lambda}$Be.
The charge symmetry
 breaking potential is not included in $^7_{\Lambda}$Be.
 The level energies are measured from
 the particle breakup threshold.}
\end{center}
\end{figure*}

In Fig.2 to Fig.4
 and Table \ref{tab:halo_energy_rms_A=7},
we show the level structures of A=7 hypernuclei calculated without
the CSB interaction.
In each figure, hypernuclear levels are shown in
four columns in order to show separately the effects of
even- and odd-state $\Lambda -N$ interactions and also the SLS and
ALS interactions.
Even if the CSB interactions are switched on, their small contributions
do not alter the features of these figures.
At first glance, the obtained $\Lambda$ states become less bound by 1 MeV
in the order of $^7_\Lambda$He, $^7_\Lambda$Li$^*$ and $^7_\Lambda$Be,
because the repulsive Coulomb-force contributions increase in this order.
In these figures,
the calculated energy spectra of low-lying states of core nuclei,
$^6$He, $^6$Li$^*$  and $^6$Be are also drawn in order to
demonstrate the $\Lambda$-binding effects.
Here, $^6$He and $^6$Li$^*$ are  nucleon-bound states,
and the $N\alpha$ and $NN\alpha$ interactions are adjusted so as
to reproduce the observed energy spectra.
On the other hand, $^6$Be is an nuclear-unbound system.
In order to extract the $B_{\Lambda}$ value in
$^7_{\Lambda}$Be,
it is needed to subtract the total energy of the lowest $^6$Be
resonant state from the calculated ground-state energy of $^7_\Lambda$Be.
The energy positions of resonant states are determined by
the real scaling method \cite{real}:
The obtained lowest state in $^6$Be is a $0^+$ broad resonance,
whose energy is 0.79 MeV.
Thus, the experimental resonant energy 1.37 MeV cannot be reproduced,
when the $\alpha N$, $NN$ and $\alpha NN$ interactions are adopted
so as to reproduce the bound-state energies of $^6$He and $^6$Li$^*$.

\begin{figure*}[htb]
\begin{minipage}{0.35\linewidth}
\scalebox{0.35}
{\includegraphics{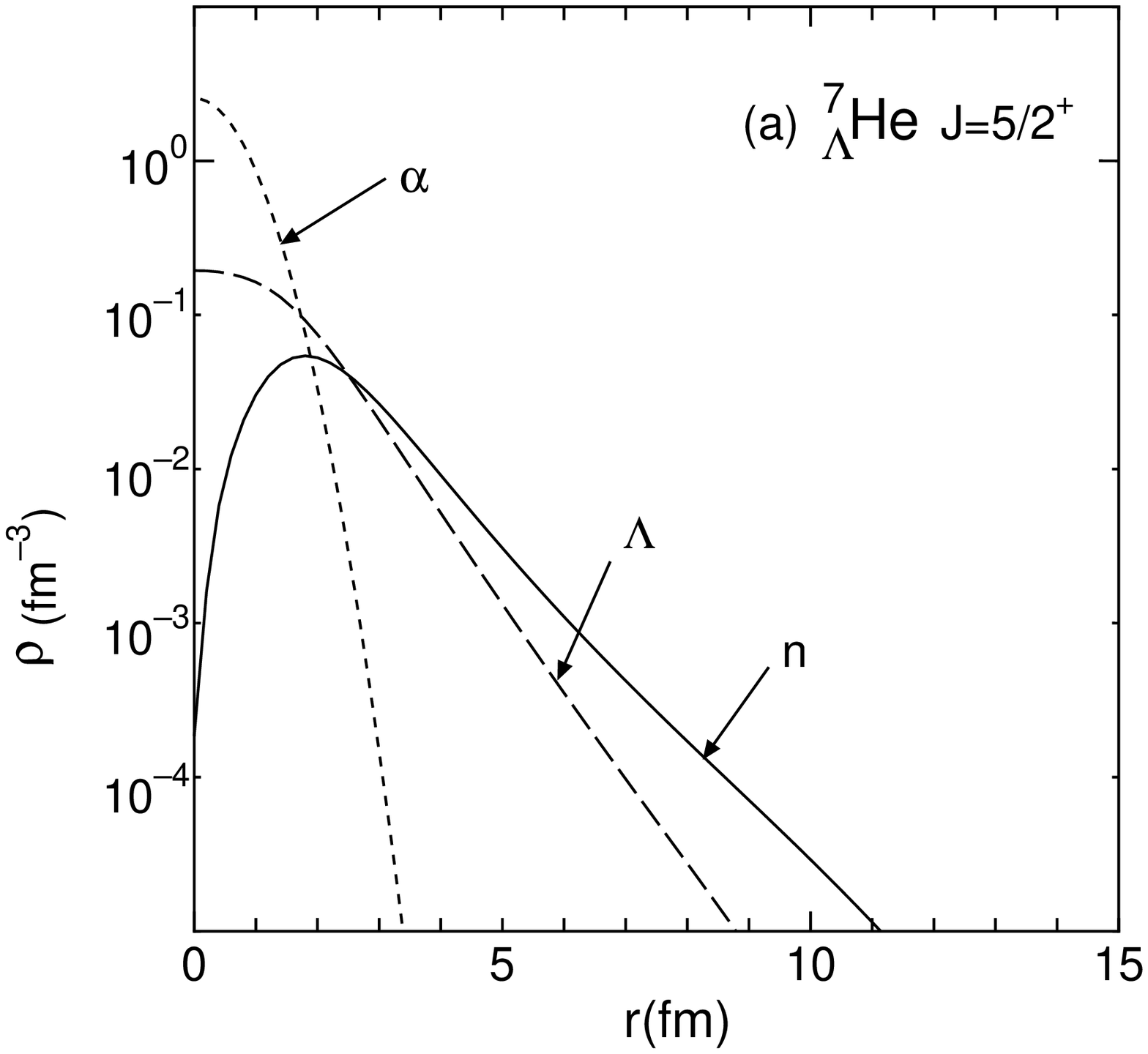}}
\end{minipage}
\begin{minipage}{0.31\linewidth}
\scalebox{0.35}
{\includegraphics{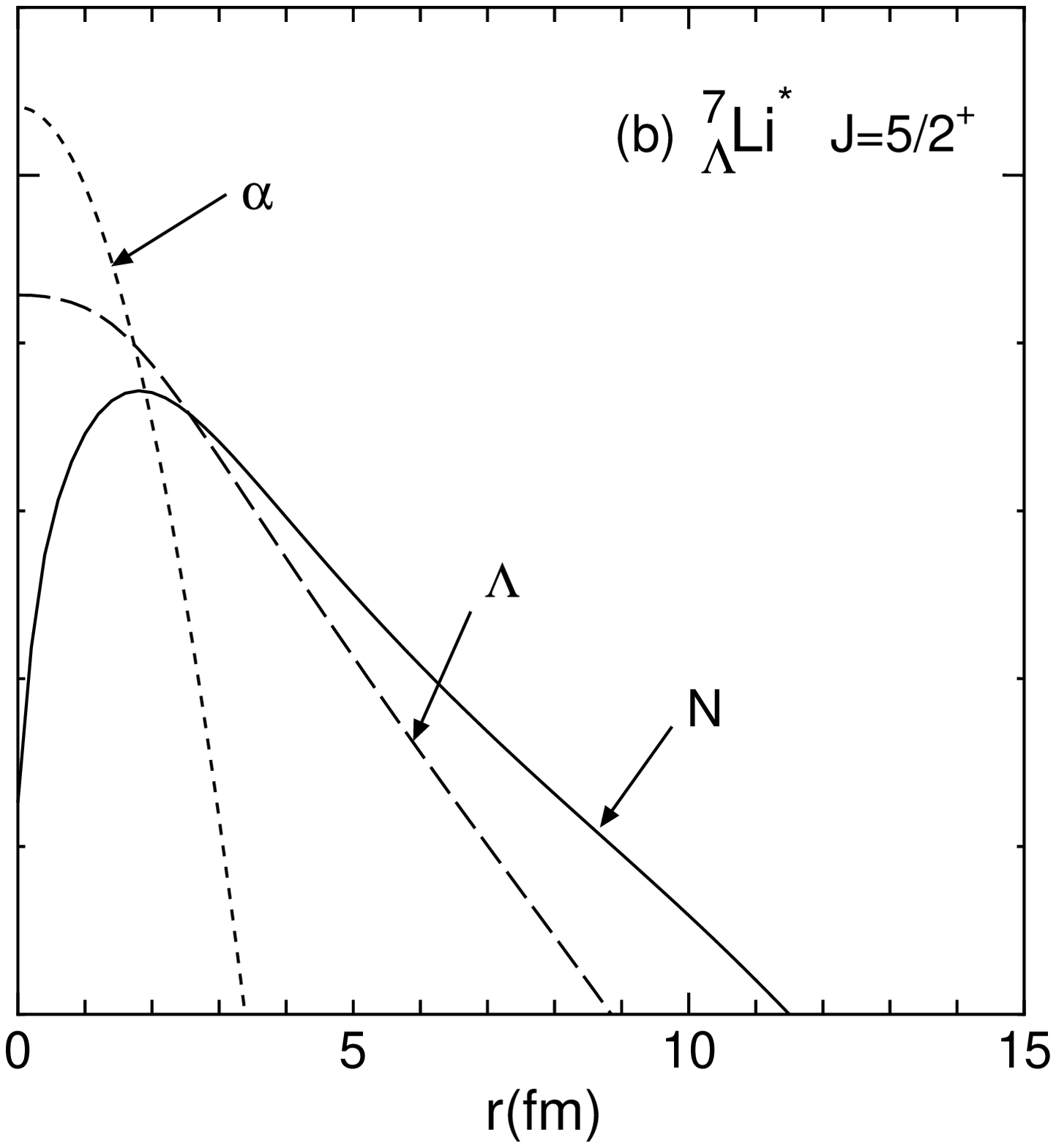}}
\end{minipage}
\begin{minipage}{0.31\linewidth}
\scalebox{0.35}
{\includegraphics{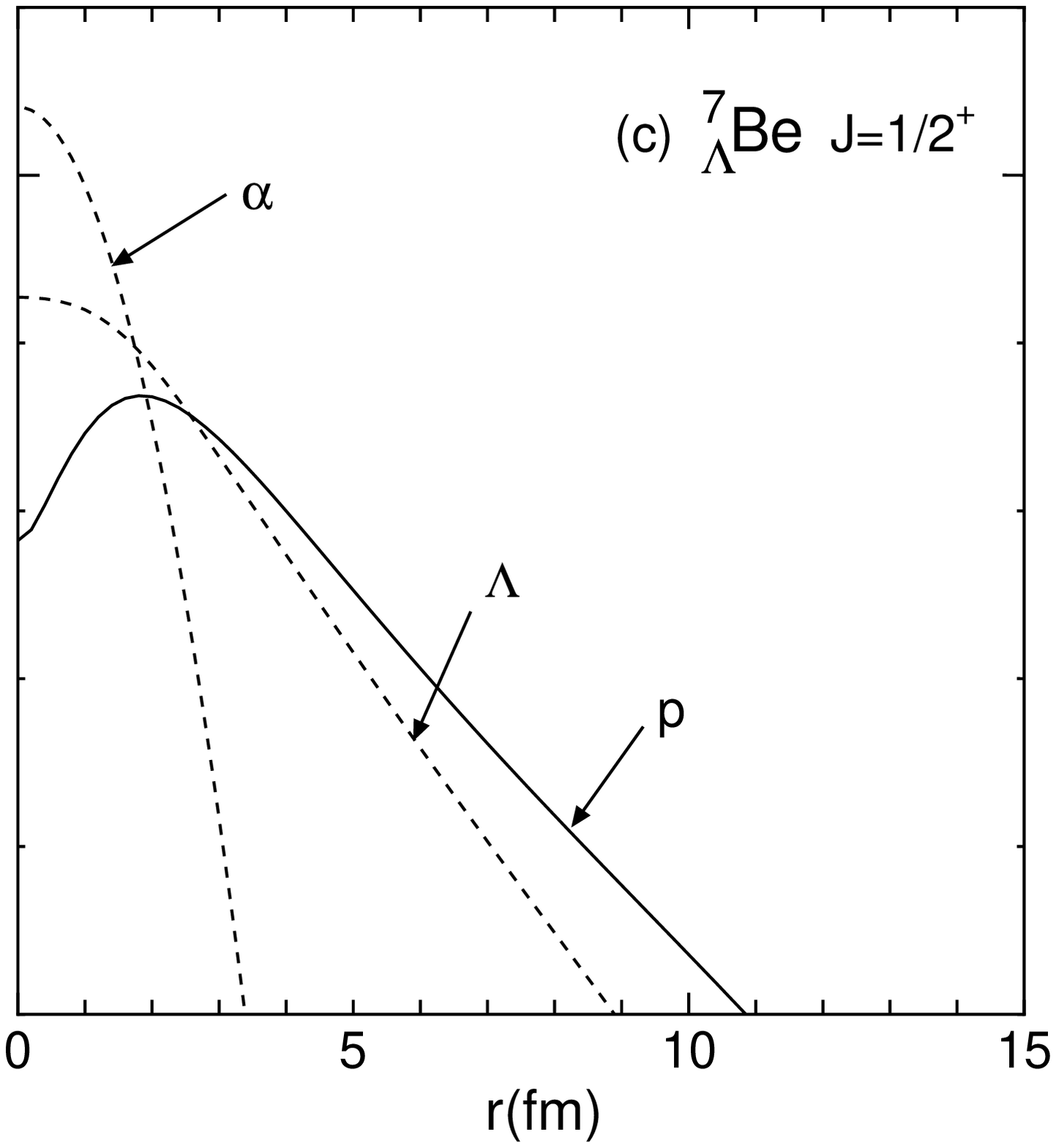}}
\end{minipage}
\caption{Calculated density distribution of
$\alpha$, $\Lambda$ and a valence nucleons
for (a)$^7_{\Lambda}$He, (b)$^7_{\Lambda}$Li$^*$ and
(c)$^7_{\Lambda}$Be without charge symmetry breaking potential.}
\label{fig:li7den}
\end{figure*}

It is particularly interesting to see the gluelike role of the
$\Lambda$ particle in $A=7$ hypernuclear systems.
Though the ground state of $^6$Be is unbound,
the $\Lambda$ participation leads to a bound state
below the lowest $^5_{\Lambda}$He$+p+p$ threshold,
the binding energy of which is about 1.3 MeV.
The ground states of the core nuclei $^6$He and $^6$Li$^*$ are
weakly bound by 1.02 and 0.12 MeV below the $\alpha +N+N$ threshold.
Owing to an additional $\Lambda$ particle,
those of $^7_{\Lambda}$He and $^7_{\Lambda}$Li$^*$ become rather deeply
bound by about 2 $\sim$ 3 MeV below the respective lowest thresholds.
It should be noted, here, that the calculated values of $B_\Lambda$
of $^7_{\Lambda}$Li$^*$ and $^7_{\Lambda}$Be are in good agreement
with the experimental values, as shown in Table \ref{tab:halo_energy_rms_A=7}.
%
The $5/2^+$ and $3/2^+$ excited states in $^7_{\Lambda}$Li$^*$ are
predicted to be in weakly bound states  with the respect to
the $^6_{\Lambda}$He$+p$ threshold.
On the other hand, the corresponding states in $^7_{\Lambda}$He are in
deeper bound states by about 1.3 MeV with respect to the
$^6_{\Lambda}$He$+n$ threshold.
This difference is because the $\alpha p$ Coulomb repulsion in
the former is not active in the latter.

In the past calculation \cite{Hiyama96}, the uppermost bound states in
$^7_{\Lambda}$He, $^7_{\Lambda}$Li$^*$ and $^7_{\Lambda}$Be
were $5/2^+$, $3/2^+$ and $1/2^+$ states, respectively.
These states are  very weakly bound structures, and
exhibit halo or skin structures having long tails in
density distributions of valence nucleons.
In comparison with these calculations, performed in the limited
three-body model space
($^5_{\Lambda}{\rm He}+N+N$), all states in $A=7$ systems become
deeper bound in the present four-body model.
This tendency is reasonable because in the present calculations
the excitation effects of a $\Lambda$ particle are fully taken into
account in the treatment with use of the $\Lambda N$ effective
interactions chosen consistently with the four-body model space.
It is instructive to compare the tail behavior of the density
distributions of valence nucleons in the four-body model with
those in the three-body model. We derive here the nucleon density
distributions of $5/2^+$ states in $^7_{\Lambda}$He and
$^7_{\Lambda}$Li and that of $1/2^+$ state in $^7_{\Lambda}$Be
using the two models.

\begin{table}
\begin{center}
\caption{
Calculated energies of the low-lying states of
(a) $^7_\Lambda$He, (b)$^7_\Lambda$Li$^*$, and
(c) $^7_{\Lambda}$Be without the charge symmetry breaking potential,
together
  with those of the corresponding states of $^6$He, $^6$Li$^*$,
  and $^6$Be, respectively.
$E$ stands for the total interaction energy among constituent particles.
The energies in the parentheses are measured from the corresponding
lowest particle-decay thresholds $^6_\Lambda$He $+$ $N$
for $^7_{\Lambda}$He and $^7_{\Lambda}$Li$^*$ and
$^5_{\Lambda}$He$+p+p$ for $^7_{\Lambda}$Be.
The calculated r.m.s. distances,
$\bar{r}_{\alpha-N}$, $\bar{r}_{\alpha-\Lambda}$
 are also listed for the bound state.}
\label{tab:halo_energy_rms_A=7}
\begin{tabular}{cccccccccccc}
\hline \hline  &&&&&&\\[-3mm] && &(a) \\
            &\multicolumn{2}{c}{$^{6}$He($\alpha nn$)} & \hspace{5mm}
   & \multicolumn{3}{c}{$^{7}_{\Lambda}$He$(\alpha nn\Lambda)$} \\  $J^{\pi}$  &$0^+$  &\quad  $2^+$  & & $1/2^+$  & $3/2^+$  &$5/2^+$ \\ \hline %
$E$	(MeV) & $-1.02$  &\quad $0.82$ &
& $-6.39$   & $-4.73$ &$-4.65$ \\
$E^{\rm exp}$(MeV) &$-0.98$ &\quad $0.83$  &    &   & &            \\
&	&         & & $(-3.10)$ & $(-1.44)$ &$(-1.34)$ \\
$B_{\Lambda}$(MeV) &     &  & & $5.36$  &$3.70$ &$3.62$    \\
$B_{\Lambda}^{\rm exp}$  (MeV)	&  & &   & &   &        \\
$\bar{r}_{\alpha-n}$(fm) &$4.27$ & & &$3.66$ &3.80 & $3.83$  \\
$\bar{r}_{\alpha-\Lambda}$(fm) & & & &$2.81$ &$2.79$ &$2.78$
 \\ && &(b) \\
            &\multicolumn{2}{c}{$^{6}$Li($\alpha np)$} & \hspace{5mm}
   & \multicolumn{3}{c}{$^{7}_{\Lambda}$Li($\alpha np\Lambda)$} \\  $J^{\pi}$  &$0^+$  & $2^+$  & & $1/2^+$  & $3/2^+$  &$5/2^+$ \\ \hline %
$E$	(MeV) & $-0.12$  &\quad $1.77$ &
& $-5.40$   & $-3.75$ &$-3.66$ \\
$E^{\rm exp}$(MeV) &$-0.14$ &\quad $1.67$  &    &   & &            \\
&	&         & & $(-2.11)$ & $(-0.46)$ &$(-0.37)$ \\
$B_{\Lambda}$(MeV) &     &  & & $5.28$  &$3.63$ &$3.54$    \\
$B_{\Lambda}^{\rm exp}$  (MeV)	&  & &   &5.26 &   &        \\
$\bar{r}_{\alpha-N}$(fm) &$4.73$  & & &$3.74$ &$3.92$ &$3.96$  \\
$\bar{r}_{\alpha-\Lambda}$(fm) & && &$2.82$ &$2.80$ &$2.80$    \\  && &(c) \\
            &\multicolumn{2}{c}{$^{6}$Be$(\alpha pp)$} & \hspace{5mm}
   & \multicolumn{3}{c}{$^{7}_{\Lambda}$Be $(\alpha pp \Lambda)$} \\  $J^{\pi}$  &$0^+$  & $2^+$  & & $1/2^+$  & $3/2^+$  &$5/2^+$ \\ \hline %
$E$	(MeV) &$0.79$  & &
& $-4.42$   &  & \\
$E^{\rm exp}$(MeV) &$1.54$ &$2.93$  &    &   & &            \\
&	&         & & $(-1.30)$ & & \\
$B_{\Lambda}$(MeV) &     &  & & $5.21$  & &    \\
$B_{\Lambda}^{\rm exp}$  (MeV)	&  & &   &5.16 &   &        \\
$\bar{r}_{\alpha-p}$(fm) & & & &$3.84$ & &   \\
$\bar{r}_{\alpha-\Lambda}$(fm) & & & &$2.83$ & & \\  \hline \hline
\end{tabular}
\end{center}
\end{table}

In Table \ref{tab:halo_energy_rms_A=7},
we list the calculated values of the rms radii
between $\alpha$ and $N$ ($\bar{r}_{\alpha - N}$)  and
those between $\alpha$ and $\Lambda$ ($\bar{r}_{\alpha - \Lambda}$)
in our four-body models of $^7_{\Lambda}$He,
$^7_{\Lambda}$Li$^*$ and $^7_{\Lambda}$Be.
As shown here, the values of $\bar{r}_{\alpha -n}$ in these systems
are larger than those of $\bar{r}_{\alpha -\Lambda}$, indicating
that the distributions of valence nucleons are of longer-ranged tail
than those of $\Lambda$'s in the respective systems.
However, all r.m.s radii in the four-body models are shorter than those
in the three-body models \cite{Hiyama96}, that is
the four-body binding energies in the present model are larger
than the three-body ones in the previous model.
This means that the distributions of nucleons and $\Lambda$
around $\alpha$ obtained in the four-body models are more compact
than those in the three-body models.

In order to see the structures of these systems visually,
in Fig.5 we draw the
density distributions of $\Lambda$ (dashed curve) and valence neutrons
(solid curve) of the $5/2^+$ states in $^7_{\Lambda}$He and
$^7_{\Lambda}$Li$^*$ and of the $1/2^+$ state in $^7_{\Lambda}$Be.
For comparison, here, also a single-nucleon density in the $\alpha$ core
is shown by  the dotted curve.
In each case, the density distribution of the $\Lambda$ has
a shorter-ranged tail than that of the two valence nucleons,
but is extended significantly far away from the $\alpha$ core.
This structure can be nicely imaged as
three layers of matter distribution composed of
an $\alpha$ core, a $\Lambda$ skin and a neutron (proton) halo.
Here, the proton-density distribution in the $5/2^+$ state of
$^7_{\Lambda}$Li$^*$ has a particularly longer tail than those in the
others due to the very weak binding of the halo proton from the
lowest $^6_{\Lambda}$He$+p$ threshold.

It is considered that the $3/2^+$-$5/2^+$ spin-doublet states in
$^7_{\Lambda}$He and $^7_{\Lambda}$Li$^*$ give valuable information
about the underlying spin-dependence of the $\Lambda N$ interaction.
Let us investigate these states straightforwardly with use of the
$\Lambda N$ interaction determined in the analysis for the $T=0$
spin-doublet states in $^7_{\Lambda}$Li.
The results for $^7_\Lambda$He and $^7_\Lambda$Li$^*$
are displayed in Fig.2 and Fig.3, respectively.
Because their features are not different from each other,
here we pick up the former case.

\begin{figure*}[htb]
\begin{center}
\epsfig{file=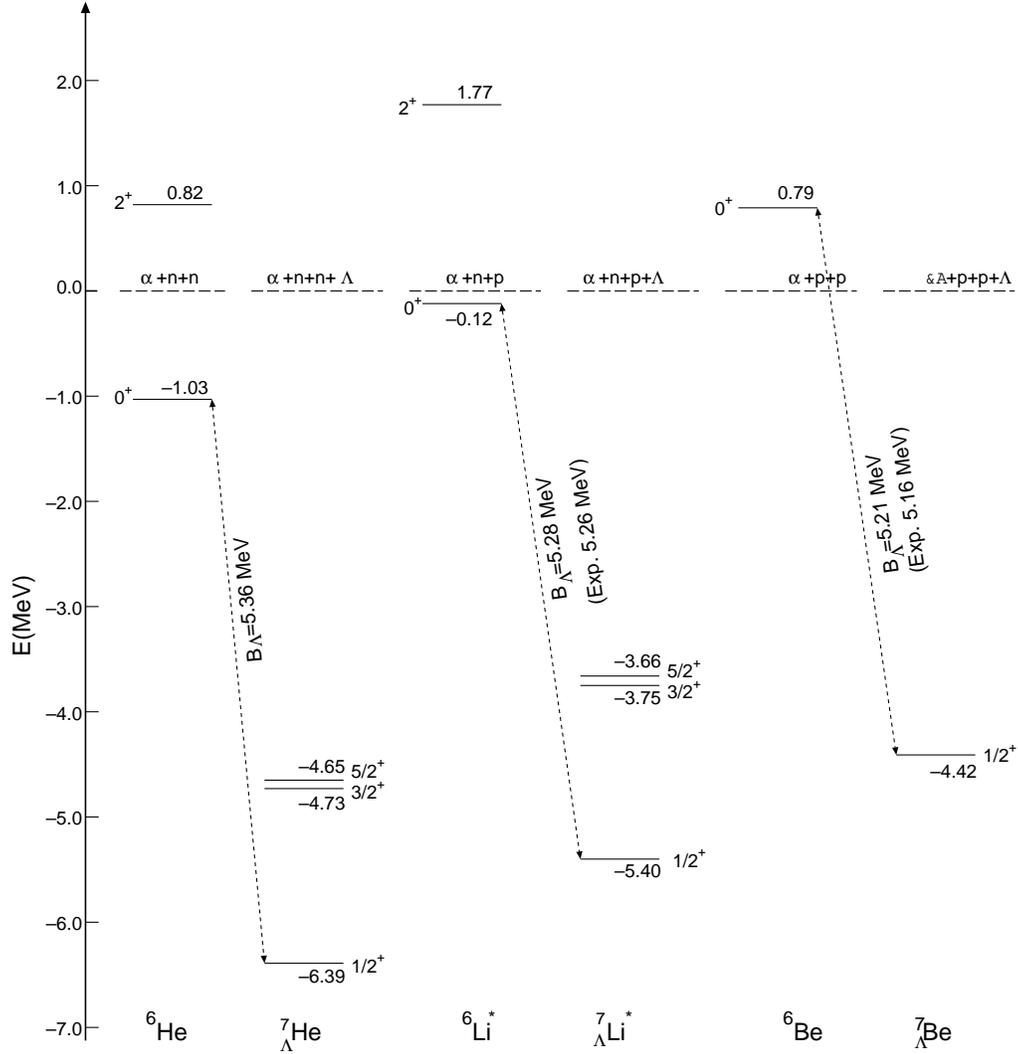,scale=0.35}
\caption{The calculated energy levels of $^6$He, $^7_{\Lambda}$He,
$^6$Li$^*$, $^7_{\Lambda}$Li, $^6$Be and
$^7_{\Lambda}$Be with spin-spin and spin-orbit $\Lambda N$ interactions.
The charge symmetry breaking potential is not included in the
calculated energies of $A=7$ hypernuclei.
The energies are measured from the particle breakup threshold.}
\end{center}
\end{figure*}

Then, let us remark how the energies of
the $3/2^+$-$5/2^+$ spin-doublet states are changed
by adding the components of  $\Lambda N$ interaction successively.
We see that the resultant
energy splitting of $5/2^+$-$3/2^+$ states in
$^7_\Lambda$He is given as about 0.1 MeV,
being the combined contributions from the spin-spin, SLS and ALS
interactions as explained below.
We can see the same tendency in $^7_{\Lambda}$Li$^*$ in Fig.3.

It should be noted here that the splitting energies of the $T=1$
$3/2^+$-$5/2^+$ states are much smaller than  those of
the $T=0$ $1/2^+$-$3/2^+$ and $5/2^+$-$7/2^+$ doublet states in
$^7_{\Lambda}$Li given in Ref.\cite{Hiyama96}.
To understand the reason for the difference between the
$T=1$ and $T=0$ doublet splittings, first we remark
that the spin-isospin structure of $NN \Lambda$ system on the
$\alpha$ core is $[(NN)_{sTT_z}\Lambda]_S$ (cf.Eq.(2.3)).
In the case of $T=1$ states,
the corresponding $nn$ pair is in spin-singlet states
($s=0$, spin antiparallel), while
in $^7_{\Lambda}$Li $(T=0$)
the $np$ pair outside the $\alpha$ core is in a spin-triplet state
($s=1$, spin-parallel).
In general the numbers of
$\Lambda N$ triplet and singlet bonds are different between
the $J_>$ and $J_<$ partner states.
Thus difference in spin-value of $(NN)_{s=1 \:{\rm or}\: 0}$
leads to the different
contributions of the $\Lambda N$ spin-spin interactions
to the doublet splittings.
Let us see in more  detail how
the $\Lambda N$ spin-spin interactions contribute to
the $3/2^+$-$5/2^+$ splitting in $^7_\Lambda$He($T=1$).
Both doublet states are composed of
the $L=2$ $(nn)_{s=0,T=1}$ pair in the spin-singlet
state coupled to the $s$-state $\Lambda$.
As mentioned above, the situation is notably different
from that of the $5/2^+$-$7/2^+$ doublet in
$^7_{\Lambda}$Li $(T=0$) which is based on the
$[L=2$ $(pn)_{s=1,T=0}]_{J=3^+}$ core state and therefore
the $J_>=$ $7/2^+$ partner is characterized by the
spin-stretched configuration. In contrast to the $T=0$ case,
both of the $J_<=$ $3/2^+$ state and the $J_>=$ $5/2^+$
state in $^7_{\Lambda}$He $(T=1$) include $\Lambda N$
spin-singlet and spin-triplet states. However, we find
that the contribution of the $\Lambda N$ spin-singlet
state is negligbly small in the $J_>=$ $5/2^+$ state.
As a result the even-state spin-spin part of the
$\Lambda N$ interaction gives rise to the splitting
energy of about 0.31 MeV (See "even" column.).
In addition, when  the odd-state interaction
is switched on,
the energy splitting is reduced to be about 0.13 (See "+odd" column.) MeV.
The major reason for this reduction is because
$V_{\Lambda N}^{(^1O)}$ is more repulsive than
$V_{\Lambda N}^{(^3O)}$, and therefore the $3/2^+$ state
including  $\Lambda N$ spin-singlet component is pushed up
more than the $5/2^+$ state.

\begin{figure*}[htb]
\begin{center}
\epsfig{file=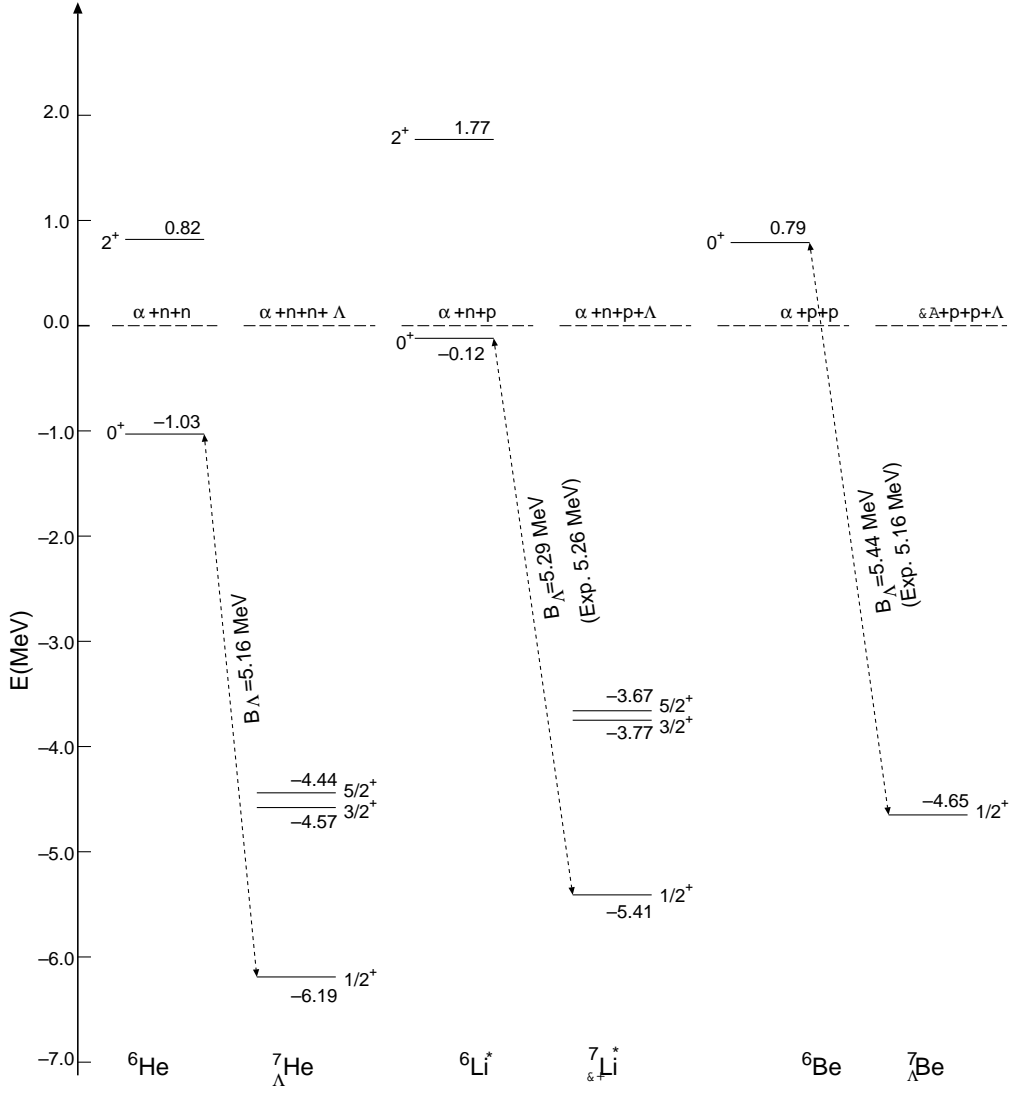,scale=0.35}
\caption{The calculated energy levels of $^6$He, $^7_{\Lambda}$He,
$^6$Li$^*$, $^7_{\Lambda}$Li, $^6$Be and
$^7_{\Lambda}$Be with spin-spin and spin-orbit $\Lambda N$ interactions.
The charge symmetry breaking potential is  included in the
calculated energies of $A=7$ hypernuclei.
The energies are measured from the particle breakup threshold.}
\end{center}
\end{figure*}

Moreover, we continue to add SLS and ALS contributions to the $3/2^+$ and
$5/2^+$ doublet states.
As shown in Fig.2, the SLS works attractively for the $5/2^+$ state,
because the contribution of the
$\Lambda N$ spin-triplet state is dominated in this state.
On the other hand, the ALS works significantly in the $3/2^+$ state,
because the ALS acts between spin=0 and 1 $\Lambda N$ states.
However, the ALS does not efficiently work in the
$5/2^+$ state, because the spin-singlet component is small in this state.
%
As a result, the energy splitting of $5/2^+$-$3/2^+$ states
including both spin-spin and spin-orbit terms in $^7_\Lambda$He
leads to 0.08 MeV.
We can see the same tendency in $^7_{\Lambda}$Li$^*$ and the resultant
splitting is 0.09 MeV as shown in Fig.3.
If the experimental energy resolution becomes good enough to discuss the present splitting energy,
we would have a chance of getting  information about the spin-dependent
parts of
the $\Lambda N$ interaction.

There still remain certain effects of
the $\Lambda N$ tensor interaction on the doublet splittings.
In this paper for the $T=1$ isotriplet states $(A=7$),
however, we apply the prescription adopted in the
analysis of the
$T=0$ $^7_{\Lambda}$Li states \cite{Hiyama06} and
therefore we do not include the tensor component.
Here we note that the $\Lambda N-\Lambda N$ tensor contribution
is small compared to the spin-spin interaction,
however another tensor effect comes from the $\Lambda N-\Sigma N$ coupling.
In fact, accounting for 
$\Sigma-\Lambda$ coupling by modifying the $\Lambda N$
interaction alters its effect on doublet splitting, and hence 
introduces an uncertainty in the calculation. According to the 
$\Sigma$-mixing studied within the shell model \cite{Millener08}, 
the energy shifts amount to several tens of keV in some of the 
$T= 0$ states of $^7_{\Lambda}$Li.
The cluster model estimates for such effect will be
discussed in the next stage.


\section{Charge Symmetry breaking effects}

\subsection{CSB effects in $A=7$ four-body models}

Let's focus on the ground states in $^7_{\Lambda}$He
and $^7_{\Lambda}$Be and the $T=1$ $1/2^+$ state in $^7_{\Lambda}$Li,
which are the members of the iso-triplet.
The CSB effect has to be reflected also in their binding energies
in the same way as in the $T=1/2$ iso-doublet members
$^4_{\Lambda}$H and $^4_{\Lambda}$He.


As explained in sec.\ref{sec:CSB}, we introduce the phenomenological CSB potential
with the central-force component only.
The CS part of the two-body $\Lambda N$ interaction is fixed
to reproduce the averaged energy spectra of $^4_{\Lambda}$H
and $^4_{\Lambda}$He,
and then the CSB part is adjusted so as to reproduce simultaneously
the energy levels of these hypernuclei.
The spin-spin part of the CSB can be determined by performing
this adjusting procedures both for the $0^+$ and $1^+$ states.


First, in Fig. 6, we show the energy spectra of $A=7$ hypernuclei
without the CSB interaction.
The ground-state energy of $^7_{\Lambda}$He is $-6.39$ MeV
with the respect to the
$\alpha +n+n+\Lambda$ four-body breakup threshold.
With increase of the proton numbers, the Coulomb repulsion
becomes more and more effective as going from
$^7_{\Lambda}$Li$^*$ to $^7_{\Lambda}$Be.
Recently in KEK-E419 experiment \cite{Tamura00}, they produced the
$T=1$ $1/2^+$ state of $^7_\Lambda$Li.
The observed value of $B_{\Lambda}$=5.26 MeV is
in good agreement with our calculated value 5.28 MeV.
In the case of $^7_{\Lambda}$Be, there are the old emulsion data
giving $B_{\Lambda}$=5.16 MeV.
This value should be compared with our obtained value 5.21 MeV.
Then,
the $B_{\Lambda}$ value in the ground $1/2^+$ state of
$^7_{\Lambda}$He is predicted to be 5.36 MeV
without taking the CSB effect into account.

Next, let's consider the CSB effects in $A=7$ iso-triplet hypernuclei.
In Fig. 7, we show the energy spectra of those hypernuclei
calculated with the CSB interaction switched on.
In the $^7_{\Lambda}$Li case,
the CSB interaction brings about almost no contribution to
the $\Lambda$ binding energies, because there is one proton
and one neutron outside the $\alpha$ core and the
$\Lambda n$ and $\Lambda p$ CSB interactions cancel with each other.
On the other hand, the CSB interaction works repulsively (+0.20 MeV)
and attractively ($-0.20$ MeV)
in the $^7_{\Lambda}$He and $^7_{\Lambda}$Be cases, respectively.
Therefore, our result indicates that if the experimental
energy resolution is as good enough as less than 0.2 MeV,
the CSB effect could be observed in these cases.
It should be noted here that only the even-state part of our CSB interaction
is taken into account in consistent with the observed
binding energies of $^4_{\Lambda}$H and $^4_{\Lambda}$He.

In the $^7_{\Lambda}$Be case, the $\Lambda$ energy becomes
 more bound by 0.2 MeV
due to the attractive CBS interaction between the $\Lambda$ and two protons,
that is $B_{\Lambda}=5.44$ MeV.
The experimental $B_\Lambda$ value is found to be reproduced
without the CSB effect and the inclusion of
the CSB contribution goes unfavorably.
In order to reproduce the binding energy of $^7_{\Lambda}$Be,
the CSB interaction seems to be vanishing or even of opposite
sign from that in the $A=4$ system.
There still remains a problem in our treatment for the $^7_{\Lambda}$Be system:
The calculated value 0.79 MeV of the lowest resonance energy of the $^6$Be
is not in agreement  with the experimental value 1.37 MeV.
When the attractive $\alpha pp$ interaction is switched off,
the $^6$Be$(0^+)$ resonance energy becomes 1.18 MeV, which is
still a bit lower than the observed value.
This change of the calculated resonance energy from 0.79 to 1.18 MeV
makes the $B_\Lambda$ value smaller by only 30 keV.
Thus, the change of the $B_\Lambda$ value is considered to be so small,
even if the $\alpha pp$ interaction is adjusted so as to just reproduce
the value 1.37 MeV.

In the $^7_{\Lambda}$He case,
the CSB interaction between the $\Lambda$ and two valence neutrons
works repulsively and the ground-state binding energy becomes
$B_{\Lambda}=5.16$ MeV, less bound by 0.2 MeV, than the value without
the CSB effect.
Though there is no data for $^7_{\Lambda}$He at present,
the $B_{\Lambda}$ of $^7_{\Lambda}$He will be obtained soon
by the $(e,e'K^+)$ reaction experiment done at JLAB.
It is interesting to know whether or not the CSB effect in
$^7_{\Lambda}$He is consistent with the emulsion data for
$B_{\Lambda}(^7_{\Lambda}$Be).


\subsection{CSB effects in $A=8$ cluster models}

Let us study another sets of two
mirror hypernuclei, $^8_{\Lambda}$Li and $^8_{\Lambda}$Be,
in the $p$-shell region within the framework of the $\alpha t \Lambda$ and
$\alpha ^3{\rm He} \Lambda$ cluster models.

The experimental values of $B_\Lambda$ from the emulsion data
are $6.80\pm 0.03$ MeV and $6.84\pm 0.05$ MeV for
$^8_\Lambda$Li and $^8_\Lambda$Be, respectively.
Thus, the energy difference $\Delta B_\Lambda^{(8)}=
B_\Lambda(^8_\Lambda$Be) $-B_\Lambda(^8_\Lambda$Li) is 0.04 MeV,
which is much smaller than the
experimental value $\Delta B_\Lambda^{(4)}=
B_\Lambda(^4_\Lambda$He) $-B_\Lambda(^4_\Lambda$H)=0.35 MeV.
It has been pointed out \cite{Gibson95} that
due to the strong Coulomb force in $A=8$ hypernuclei,
$\Delta B_\Lambda^{(8)}$ seems small and hence charge symmetry
breaking effect seems small.
It is interesting to see how much
$\Delta B_\Lambda^{(8)}$  is obtained in an actual
microscopic calculation by introducing the phenomenological
CSB interaction.

In our previous work \cite{Hiyama96},
 the cluster model calculations were performed
for these hypernuclei with use of the charge symmetric
$\alpha$-$t(^3{\rm He})$,
$\Lambda$-$\alpha$ and $\Lambda$-$t(^3{\rm He})$ interactions adjusted so as to
reproduce the experimental value 6.80 MeV for $^8_\Lambda$Li.
Then, the obtained value of $B_\Lambda$ was 6.72 MeV for $^8_\Lambda$Be,
where the difference from the value for $^8_\Lambda$Li was only due to
the difference of the Coulomb-force contributions.

In order to see the effect of the CSB interaction, we repeated
the energy level calculations employing the present interactions
given in the section \ref{interaction}.
When only the CS parts of $\Lambda N$ interactions are used,
the calculated values of $B_\Lambda(^8_\Lambda$Li) and
$B_\Lambda(^8_\Lambda$Be) are 6.80 and 6.84 MeV, respectively.
Here these CS parts are slightly modified from that in Ref.\cite{Hiyama02}
so as to reproduce well the experimental value of
$B_\Lambda (^8_\Lambda$Li) finally.
Switching on the CSB parts,
the calculated values of $B_\Lambda$ become 6.74 and 6.90 MeV for
$^8_\Lambda$Li and $^8_\Lambda$Be, respectively,
Then, the calculated value of $\Delta B_\Lambda^{(8)}=
B_\Lambda(^8_\Lambda$Be) $-B_\Lambda(^8_\Lambda$Li) is 0.16 MeV.
Thus, the use of the CSB interaction determined in the $A=4$ systems
leads to a larger value of $\Delta B_\Lambda^{(8)}$ in comparison with
the experimental value of 0.04 MeV. In order to reproduce
the experimental value of $\Delta B_\Lambda^{(8)}$, here,
let us have a try to introduce an odd-state CSB interaction
phenomenologically,
whose contributions in the $A=4$ systems are negligible:
We found that
the experimental values of $B_\Lambda$ for $^8_\Lambda$Li and
$^8_\Lambda$Be can be reproduced by adding a rather long-ranged
odd-state interaction with the opposite sign of the
even state CSB interaction described in Eq.(3.5).
The $B_{\Lambda}$ values of $^8_{\Lambda}$Li and $^8_{\Lambda}$Be
calculated with both even-state and odd-state CSB interactions
are 6.81 MeV and 6.83 MeV, respectively, which are
in good agreement with the data.

The present framework for the $A=8$ iso-multiplet systems
has a sort of limitation in the sense that the $t(^3{\rm He})$ cluster
is assumed to have 3 nucleons of the same size of those in $\alpha$.
However, the results for both systems of $A=7$ and 8 are qualitatively
 consistent with
each other, and the odd state of the CSB interaction are found to
have an opposite sign of the even state CSB interaction
determined at $A=4$ hypernuclei.

In the near future, we expect to have the observed $B_{\Lambda}$ of
$^7_{\Lambda}$He from the $(e,e'K^+)$ reaction experiment
done at JLAB. On the basis of the coming data, it might be possible
to get information on the odd-state CBS interactions.
Another example to clarify the even- and odd-state CSB interactions
is to study $^{10}_{\Lambda}$Be with an $\alpha \alpha N \Lambda$
four-body model.
This four-body calculation is in progress.
Also, we hope to observe the $B_{\Lambda}$
of this hypernucleus by $^{10}$B$(e,e'K^+)^{10}_{\Lambda}$Be
at JLAB in the future.

\section{Summary}
We have studied the structures of the $T=1$ triplet hypernuclei ($^7_{\Lambda}$He, $^7_{\Lambda}$Li and $^7_{\Lambda}$Be) within the framework of $\alpha +\Lambda+N+N$ four-body model.
In the previous paper this four-body model proved to work successfully in the detailed analysis of the $T=0$ energy levels of $^7_\Lambda$Li which are best known through the high-resolution $\gamma$-ray measurements. The present framework is also a natural extension of the previous calculations performed with the $^5_{\Lambda}$He$+N+N$ three-body model in which the $\Lambda$ particle motion was confined to form the $^5_{\Lambda}$He ground state.

 The major conclusions are summarized as follows:

(1) On the basis of  reasonable $\alpha p(n)$, $\alpha pn$,
$\alpha \Lambda$ and $N \Lambda$ interactions, which well describe the binding energies of all sub-cluster units ($\alpha pn$, $\alpha \Lambda$ and $N \Lambda$), we have made extensive and successful structure analyses for the $T=1$ states of $A=7$ iso-triplet hypernuclei.
One of the non-trivial and important outcomes is that the observed $B_\Lambda$ value of the $T=1$ $1/2^+$ state in $^7_\Lambda$Li is reproduced nicely with
the use of the $\alpha \Lambda$ and $\Lambda N$ interactions determined in $T=0$ states of $^7_\Lambda$Li.
Also the $B_\Lambda (^7_\Lambda$Be) observed in emulsion is reproduced well, though there still remains a problem that the unbound $^6$Be $0^+$ state is calculated at a bit lower position in comparison with the observed resonance energy.
The $\Lambda$ binding energy for $^7_\Lambda$He $(J=1/2^+)$, which has not been observed so far, is calculated to be around 5.16$-$5.36 MeV (with or without the CSB interaction).
This result will be tested when the result of the $^7$Li$(e,e' K^+)^7_{\Lambda}$He experiment comes from JLAB.

(2) As one of the purposes of the extended calculations, we have carefully tested whether the $3/2^+$ and $5/2^+$ spin-doublet excited states ( $s_{1/2}$ $\Lambda$ coupled to the $2^+$ excited core) are bound or not, since they were calculated previously to be just above the nucleon breakup threshold (weakly unbound) as a result of the limited three-body model of $^5_{\Lambda}$He+N+N.
It is interesting to see the gluelike role of the $\Lambda$ particle
carefully when it is added to the core nuclei having a
nucleon halo structure as concerned here.
In this paper the four-body calculation, which allows free motion of $\Lambda$, gives a clear prediction that the excited spin-doublet states in $^7_{\Lambda}$He ($^7_{\Lambda}$Li) become bound, respectively, at 1.3 MeV (0.3 MeV) below than the lowest nucleon-breakup threshold $^6_{\Lambda}{\rm He}+n$ ($^6_{\Lambda}{\rm He}+p$). The energy splitting between these $T=1$ doublet states comes from the spin-spin and spin-orbit interactions, which is calculated to be around 
0.1 MeV. If any coincidence experiment is available and the energy resolution is good enough to resolve the 0.1 MeV splitting, one would have a chance of extracting information on the spin-dependent interactions.
In $^7_{\Lambda}$Be, however, we do not expect to get the corresponding bound excited states.

(3) It is interesting to get the three-layer structure of the matter distributions in the $T=1$ iso-triplet hypernuclear states, which consist of the $\Lambda$ particle coupled to the nuclear core having neutron or proton halo. The typical numbers of the r.m.s. radii for the $^7_{\Lambda}$He(J=$5/2^+$),
$^7_{\Lambda}$Li$^*$(J=$5/2^+$) and $^7_{\Lambda}$Be(J=$1/2^+$) states are calculated to be $\bar{r}_{\alpha}=1.4$ fm for innermost $\alpha$, $\bar{r}_{\alpha-\Lambda}=2.8$ fm for the $\Lambda$ distribution, and $\bar{r}_{\alpha-n}=3.8$ fm for the outermost valence nucleon distribution.

(4) The charge symmetry breaking effects in light $p$-shell hypernuclei have been investigated quantitatively for the first time on the basis of the phenomenological CSB interaction which describe the experimental energy difference
between $B_\Lambda(^4_\Lambda$H) and $B_\Lambda(^4_\Lambda$He).
Here we found that the inclusion of this CSB interaction gives rise to push up the $^7_\Lambda$He energy by 0.20 MeV, but it pushes down the $^7_\Lambda$Be energy by 0.20 MeV.
In $^7_\Lambda$Li$^*$, the level energies remain unchanged by adding the CSB interaction because of cancellation between contribution of
 valence proton and neutron on $\alpha$.
Comparing the calculated value of $B_\Lambda(^7_\Lambda$Be) with the emulsion
data, it seems that the CSB interaction makes the agreement worse.
In the case of $^7_\Lambda$Be, however, there remains a problem of treating
the unbound $^6$Be core within our framework.
The CSB effect is expected to appear more clearly in the coming data of
$^7_\Lambda$He, whose core nucleus $^6$He is a bound system.
Next, we have tried to explain the binding energy difference of $T=1/2$ iso-doublet $A=8$ hypernuclei ($^8_\Lambda$Li, $^8_\Lambda$Be), adopting the phenomenological three-body models of $\alpha+t+\Lambda$ and $\alpha+^3{\rm He}+\Lambda$, respectively.
The energy difference between $^8_\Lambda$Li and $^8_\Lambda$Be, obtained in
emulsion, cannot be reproduced accurately with use of our CSB interaction.
Thus, our analyses for $p$-shell hypernuclei demonstrate that the CSB
interaction determined in the $^4_\Lambda$H and $^4_\Lambda$He doublet
is not necessarily consistent with the experimental $B_\Lambda$ values
of $^7_\Lambda$Be, $^8_\Lambda$Li and $^8_\Lambda$Be in emulsion.

(5) As a trial, we have introduced the odd-state component of the CSB
interaction, which is of a longer range than the even-state one.
In order to reproduce the experimental data of $^8_\Lambda$Li and
$^8_\Lambda$Be, it is found to be necessary that the sign of the odd-state
part is opposite with respect to that of the even part.
It is likely that such an odd-state CSB interaction plays some role in
the above $A=7$ four-body systems.

It is known that the CSB are generated essentially by the mass difference
within the $\Sigma$-multiplet mixed, and the $\Lambda -\Sigma^0$ mixing in
the meson-theoretical model.
Thus, in order to get a  firm conclusion on this matter,
it is necessary to perform four-body calculation of $A=4$ $\Lambda$ hypernuclei
and $A=7$ $\Lambda$ hypernuclei taking $NNN\Lambda$ and $NNN\Sigma$ and
$\alpha \Lambda NN$ and $\alpha \Sigma NN$, respectively.
These types of calculation are in progress.

%
%
\section*{Acknowledgments}
The authors thank Professors O. Hashimoto, H. Tamura,
 B.\ F.\ Gibson and Th. A. Rijken for helpful discussions.
This work was supported by a Grants-in-Aid for
Scientific Research from Monbukagakusho of Japan(20028007, 21540288 and 
21540284).
The numerical calculations were performed on the
HITACHI SR11000 at KEK.

\end{document}